\documentclass[a4paper, 12pt]{article}
\usepackage{amssymb,amsmath,latexsym,color,graphicx,stmaryrd}
\usepackage{hyperref}

\newtheorem{dref}{Definition}[section] 
\newtheorem{theo}[dref]{Theorem} \newtheorem{prop}[dref]{Proposition}

\newtheorem{cor}[dref]{Corollary}

 \def\loc{\mathop{\rm loc}\nolimits}

\def\Tr{\mathop{\rm Tr}\nolimits}

\def\e {\mathop{\varepsilon}\nolimits}

\title{Lagrangian intersections and glancing points: typical transitions of phase in semiclassical
approximations}

\author{Ilya BOGAEVSKII$^1$ \& Michel ROULEUX$^2$}
\date{}
\begin{document}

\maketitle

\vskip-12pt
{$^1$Mech\&Math Faculty, Lomonosov Moscow State University, Russia; Scientific Research Institute for System Analysis of RAS, Russia; The University of Liverpool, UK
\\
ibogaevsk@gmail.com}
\vskip6pt
{$^2$Aix Marseille Univ, Universit\'e de Toulon, CPT, CNRS, France
\\
michel.rouleux@univ-tln.fr}

\vskip24pt

\begin{abstract}
Given a semiclassical distribution $f_h$ microlocalized on a Lagrangian manifold $\Lambda_0$, $H\in C^\infty( T^\ast {\bf R}^n)$,
  and $H=E$ a regular energy surface, we find asymptotic solutions of the
  PDE $(H(x,\widehat{p})-E) \, u_h (x,E)=f_h(x)$ in terms of the Maslov canonical operator, when the Hamilton vector field $v_H$
  fails to be transverse to $\Lambda_0$ at some points.
\end{abstract}
\medskip


\section{Introduction}
\setcounter{equation}{0}

Given $H\in C^\infty( T^\ast {\bf R}^n)$, we discuss the semiclassical approximation to the solution of the equation
\begin{equation*}
\bigl( \widehat{H} - E \bigr) u_h (x,y,E) = f_h(x), \quad
\widehat{H} = H (x,\widehat{p}), \quad \widehat{p}= -i h \partial_x
  \end{equation*}
with a right hand side $f_h$ microlocalized on a smooth Lagrangian submanifold $\Lambda_0$:
\begin{equation*}
f_h(x) = [\mathcal{K}_{\Lambda_0} a](x), \quad a: \Lambda_0 \to {\bf R}
\end{equation*}
where $a$ is an amplitude on $\Lambda_0$, and $\mathcal{K}_{\Lambda_0}$ denotes the Maslov canonical operator.
The solution of this problem is formally given by:
\begin{equation*}
  u_h(x,E) = \frac{i}{h} \int_{0}^{+\infty} v_h(x,t) \exp(i E t/h)\, dt
  \end{equation*}
via the solution of the Cauchy problem for the Schr\"odinger equation
\begin{equation}\label{v}
i h \, \partial_t v_h = \widehat{H} v_h, \quad v_h(x,0) = f_h(x).
\end{equation}
We formulate first a set of standard hypotheses, which are sufficient in the framework of formal asymptotics we are considering here.

\begin{enumerate}
\item
$E$ is non critical for $H$, i.e. $dH\neq0$ on $\Sigma_E=\{H=E\}$.

\item
  $E$ is non trapping energy level for $H$, in the sense of scattering theory.
  This allows to cut-off the integral defining $u_h(x,E)$ near $t=+\infty$ as in (\ref{1.10});

\item
``non-return condition'' \cite{MelUh} which is not necessary if we content ourselves
to evaluate microlocally $u_h(x, E)$ outside $\Lambda_0$. Otherwise this is a difficult problem
involving Sommerfeld radiation condition, see e.\,g. \cite{Ca};

\item
$\Lambda_0\cap \Sigma_E$ is compact (isotropic) submanifold, which is fulfilled if $H$ is of elliptic type.
\end{enumerate}

Let ${\bf R}^{n+1}_{x,t} = \left\{ (x,t) \right\}$ be the space-time and $T^\ast{\bf R}^{n+1}_{x,t}$ be its cotangent bundle with the coordinates
$(x,t,p,E)$ such that $p\, d x - E \, d t$ is the canonical 1-form. To define the semiclassical approximation for $v_h$ one needs:

\begin{itemize}
\item
the Lagrangian submanifold of $T^\ast{\bf R}^{n+1}$
$$\Lambda =\left\{ (x,p) = g^t(z), t>0, E=H(z) : z \in \Lambda_0  \right\}$$
where $g^t=\exp tv_H$ is the phase flow of the Hamiltonian vector field $v_H$ generated by $H$
$$\dot{x} = \partial_{p} H(x,p), \quad \dot{p}= - \partial_{x} H(x,p).$$

\item
the amplitude $b: \Lambda \to {\bf R}$ being the solution to the transport equation with an initial condition
$b|_{\Lambda_0} = a$. Then $v_h(x)$
is given by
\begin{equation*}
v_h(x,t) = [\mathcal{K}_\Lambda b](x,t) + {\cal O}(h), \quad b: \Lambda \to {\bf R}.
\end{equation*}
\end{itemize}

So we can expect the following semiclassical approximation
\begin{equation}\label{int}
u_h(x,E) =
\frac{i}{h} \int_{0}^{+\infty} [\mathcal{K}_\Lambda b](x,t) \exp(i E t/h) \, d t \,+{\cal O}(h).
\end{equation}
The relevant contributions to this integral come from $t=0$, $t=+\infty$ and the critical points $t \in (0, + \infty)$. See also \cite{KaMa}
for higher order approximation.

The Lagrangian submanifold of $T^\ast{\bf R}^{n}$
\begin{equation*}
  \Lambda_+^E=\{z \in T^*{\bf R}^n : \exists \, t\geq0, \, z \in \exp tv_{H}(\Lambda_0 \cap \Sigma_E)\}
\end{equation*}
gives the contribution of the critical points $t \in (0, + \infty)$. The following two cases are possible.

1) The Lagrangian submanifold $\Lambda_0$ is transversal to $\Sigma_E$ --- it means that $E$ is not a critical value for the restriction $H_{\Lambda_0}$.
Then $v_{H}$ is not tangent to $\Lambda_0$ and consequently to the submanifold $\Lambda_0 \cap \Sigma_E$.
Therefore $\Lambda_+^E$ is an immersed Lagrangian submanifold with boundary and we say that
$(\Lambda_0,\Lambda_+^E)$ is a {\it (Lagrangian) intersecting pair}.

2) The Lagrangian submanifold $\Lambda_0$ is tangent to $\Sigma_E$ at a point $z$ called {\it glancing} --- it means
that $z$ is a critical point of the restriction $H_{\Lambda_0}$ with the  critical value $E$.
Then $v_{H}$ is tangent to $\Lambda_0$ and $\Lambda_+^E$  may have singularities, see e.\,g. \cite{CdV}
for applications to the se\-mi\-classical context.
\\

\noindent {\it Remark}:
A model problem with the point $t=+\infty$ giving a relevant contribution into the integral \eqref{int} is investigated in \cite{BoDoTo}.

\section{Basic example of intersecting pair}
\setcounter{equation}{0}
A simple example of a Lagrangian intersecting pair is given by
$\widetilde\Lambda_0=\left\{ x = 0 \right\}$ (vertical fiber at 0)
and
\begin{equation*}
\widetilde\Lambda_+^0=\{(x,p)\in T^*{\bf R}^n: x'=0, p_n=0, x_n \ge 0\}, \
x'=(x_1, \dots, x_{n-1}), \ p'=(p_1, \dots, p_{n-1}),
\end{equation*}
the flow-out of $\widetilde\Lambda_0$ by the Hamilton vector field
with Hamiltonian $p_n$. In this case $\widehat{H}= -i h \partial_{x_n}$,
\begin{equation*}
f_h(x) = [\mathcal{K}_{\widetilde\Lambda_0} a](x) = \int_{\widetilde\Lambda_0} e^{ixp/h}a(p)\,dp = f_1 \left( \frac{x}{h} \right)
\end{equation*}
where $f_1$ decreases rapidly at infinity,
$$v_h(x,t) = f_h(x', x_n-t)$$
and
\begin{equation}\label{1.10}
  \begin{aligned}
    u_h&(x,0) = \frac{i}{h} \int_{0}^{+\infty} f_h(x',x_n-t)\,dt =\\
    &\frac{i}{h}\int_0^{+\infty}\Theta_{t_0}(t) \, f_1 \left( \frac{x'}{h},\frac{x_n-t}{h} \right) \,dt + {\cal O}(h^\infty)
  \end{aligned}
  \end{equation}
for all $x_n\leq t_0/2$, here $\Theta_{t_0}$ is a cut-off function equal to 1 on $[0,t_0]$ and $0$ near $+\infty$.

This pair of Lagrangian manifolds is actually the paradigm of intersecting pairs,
    i.\,e. in some local canonical charts $\Lambda_0 = \widetilde\Lambda_0$ and $H(x,p)= E+p_n$.
In particular
    $(\Lambda_0,\Lambda_+^E)$ are mapped onto $(\widetilde\Lambda_0,\widetilde\Lambda_+^0)$ by a canonical transformation
    sending
    $\partial\Lambda_+^E=\Lambda_0\cap\Lambda_+^E$ onto $\partial\widetilde\Lambda_+^0=\widetilde\Lambda_0\cap\widetilde\Lambda_+^0$.

    \section{Main motivation: Helmholtz equation and Bessel cylinder}
    \setcounter{equation}{0}

Most intrinsic formulae are available in the case of the physically relevant case of a Hamiltonian positively homogeneous with respect to $p$,
e.\,g. the Helmholtz operator. This provides global eikonal coordinates and avoids the microlocal reduction to $H= E+\widehat{p}_n$.

Taking into account glancing intersection amounts
to ``correct'' locally the formulae giving the phase function and the half-density in new local coordinates,
which are obtained using a normal form.

Let us consider the \textit{Bessel cylinder} $\Lambda_0\subset T^* {\bf R}^n$
  \begin{equation}
  \label{BC}
  \Lambda_0=\{x=\varphi\omega(\psi), \, p=\omega(\psi) : \varphi\in{\bf R}\}
  \end{equation}
  where $\omega(\psi)$ is a vector
  on the unit sphere in ${\bf R}^n$.

Our main motivation is the study of Bessel beams, i.\,e.
  a wave whose amplitude is described by a Bessel function of the first kind. Assume for instance we are given a non-linear
  Helmholtz equation on ${\bf R}^2$ of the type
  $$-h^2\Delta u-u=F(\e u), \quad F(0)=0, \quad F'(0)=1$$
  where $F$ is a smooth function and $\e$ is a small parameter.
  We expand $u=u_0+\e u_1+\cdots$, and find at zero order in $\e$ the equation $(-h^2\Delta-1)u_0=0$.
  Its radially symmetric solution is given by $u_0=f_h(x)$ where
  $$f_h(x)=(2\pi/h)^{1/2}J_0\bigl({|x| / h}\bigr),$$
and is microlocalized on the Bessel cylinder $\Lambda_0\subset T^* {\bf R}^2$ defined by (\ref{BC}) for $n=2$.

  At first order in $\e $ we get the Helmholtz equation
  $$(-h^2\Delta-1) u_1 =f_h, \quad H=p^2$$
  where $u_1 = u(x,1)$ in our previous notation, and
  $$
  u_h(x,1) = -J_1(|x|/h) (|x|/2 h)
  $$
  is its radially symmetric solution. We observe that all points of $\Lambda_0$ turn to be glancing
  because $\Lambda_0 \subset \Sigma_1$ and $H_{\Lambda_0} \equiv 1$.

  For the Helmholtz equation with variable coefficients, we have the following:

  \begin{prop} Let $\Lambda_0$ be the $n$-dimensional Bessel cylinder (\ref{BC}) and $H\in C^\infty(T^*{\bf R}^n)$ be homogeneous
    of degree $m$ with respect to $p$. Then $z=(x,p)\in\Lambda_0$ is a glancing point at energy $E$ iff
    \begin{equation}\label{gl2}
      \begin{aligned}
&\partial_p H(z) + \varphi\partial_x H(z) = m H\omega(\psi),\\
&\langle-\partial_x H(z), \omega(\psi) \rangle=0, \quad H(z)=E
      \end{aligned}
    \end{equation}

  \end{prop}

\noindent {\it Proof}: We complete $\omega(\psi)$ in ${\bf S}^{n-1}$ into a (direct) orthonormal basis
$\omega^\perp(\psi)=\bigl(\omega_1(\psi),\cdots,\omega_{n-1}(\psi)\bigr)$ of ${\bf R}^n$, and denote by
$\omega^\perp(\psi)\delta\psi=\omega_1(\psi)\delta\psi_1+\cdots+\omega_{n-1}(\psi)\delta\psi_{n-1}$ a section of $T{\bf S}^{n-1}$,
$\delta\psi_j\in{\bf R}$.
The tangent space $T_z\Lambda_0$ has the parametric equations
\begin{equation*}
  \delta X=\omega(\psi)\delta\varphi+\varphi\omega^\perp(\psi)\delta\psi, \
 \delta P=\omega^\perp(\psi)\delta\psi, \ (\delta\varphi,\delta\psi)\in{\bf R}^n
\end{equation*}
so $v_H\in T_z\Lambda_0$ iff there exist $(\delta\varphi,\delta\psi)$ such that
\begin{equation*}
  \partial_pH=\omega(\psi)\delta\varphi+\varphi\omega^\perp(\psi)\delta\psi-\partial_xH=\omega^\perp(\psi)\delta\psi
  \end{equation*}
Taking scalar products with $\omega(\psi),\omega^\perp(\psi)$, and using that
$(\omega(\psi),\omega^\perp(\psi))$ form a basis of ${\bf R}^n$, readily shows that relations
\begin{equation}\label{gl4}
  \partial_p H+\varphi\partial_x H=\langle\partial_p H,P(\psi)\rangle\omega(\psi), \
  \langle-\partial_x H,\omega(\psi)\rangle=0
  \end{equation}
are necessary and sufficient
for $v_H\in T_z\Lambda_0$.

We set ${\cal H}(\varphi,\psi)=H|_{\Lambda_0}$.
Then
\begin{equation*}
  \nabla{\cal H}(\varphi,\psi)=\bigl(\langle\partial_xH,\omega^\perp(\psi)\rangle, \varphi\langle\partial_xH,\omega^\perp(\psi)\rangle+
  \langle\partial_pH,\omega^\perp(\psi)\rangle\bigr)
  \end{equation*}
so (\ref{gl4}) readily gives
$\nabla{\cal H}(\varphi,\psi)=0$.
Therefore $z=(x,p)\in\Lambda_0$ is a glancing point.

Now, if $H$ is positively homogeneous of degree $m$ with respect to $p$,  using Euler identity, we get
$\delta\varphi=\langle\partial_p H,P(\psi)\rangle=mH$,
and (\ref{gl2}) holds iff
for $v_H\in T_z\Lambda_+^E\cap T_z\Lambda_0$ when $H=E$. $\clubsuit$ \\

In particular when $H(z)=\frac{|p|^m}{\rho(x)}$ is a conformal metric, with $\rho$ a smooth positive function on ${\bf R}^n$,
$z$ is a glancing point iff
\begin{equation*}
    \begin{aligned}
    &\hbox{either}: \ \varphi\neq0 \ \hbox{and} \ \nabla\rho=0, \cr
    &\hbox{or}: \ \varphi=0 \ \hbox{and} \
    \langle\nabla\rho(0),\omega(\psi)\rangle=0.
    \end{aligned}
\end{equation*}

\noindent {\it Example}: Let $n=2$, $m=1$, $H(z)=\frac{|p|}{\rho(x)}$, with
$\rho(x)=\frac{1}{2}(1+(x-x_0)^2)$. If $x_0=\varphi\omega(\psi)\neq0$, we have
$\rho^4(x_0)\det \nabla^2 (H|_{\Lambda_0})=\varphi^2$, $\rho^2(x_0)\Tr \nabla^2 (H|_{\Lambda_0})=-(1+\varphi^2)$.
Critical energy is given by $E=H(z)=\frac{|p|}{\rho(x)}$, i.\,e. $E_0=1/\rho(x_0)$.\\

\noindent {\it Remark}:
  Let $n=2$, $H(z)$ positively homogeneous of degree $m=1$.
  Away from glancing points, $\Lambda_+^E$ is parametrized near $t=0$ by $(\psi,\theta^E)$, where
$\theta^E=\theta^E(t,\varphi)$ and $\theta^E:{\bf R}^2 \to {\bf R}$
  is a submersion, such that
  $$H\bigl(X(t,\varphi,\psi),P(t,\varphi,\psi)\bigr)=E$$
When $\rho(x)=\rho(|x|)$, we have $\varphi=\varphi(E)$ on $\Lambda_+^E$,
so that we can take $\theta^E(t,\varphi)=Et$ (or equivalently $\theta^E(t,\varphi)=\varphi(E)+Et$, which is the eikonal on $\Lambda_+^E$),
and $\Lambda_+^E$ is parametrized by $(t,\psi)$.
Conversely, if $\Lambda_+^E$ is parametrized by $(t,\psi)$,
the fact that $\Lambda_+^E$ is Lagrangian implies the symmetry relation
$\langle \dot X, P_\psi\rangle=\langle \dot P, X_\psi\rangle$.
For $H=\frac{|p|}{\rho(x)}$, taking the limit $t\to 0$ in Hamilton equations gives
$\varphi\langle \nabla\rho, \omega^\perp(\psi)\rangle=0$, which shows that $\rho$ is radially symmetric.\\

\section{Eikonal coordinates and generating families in the case $m=1$}
\setcounter{equation}{0}

Let $\iota:L\to T^*{\bf R}^d_x$ be a smooth immersed Lagrangian manifold,
 the 1-form $p\,dx$ is closed on $L$, and so locally $p\,d x = d S$.
 If $p\,d x \neq 0$ then $S$ can be chosen as a (local) coordinate on $L$.
 Following \cite{DoMaNaTu}, \cite{DoNaSh} we say that $S$ is the {\it eikonal} or the {\it action}
 on $L$ which can be completed to a system of {\it eikonal coordinates} on $L$.

 Let $L=\Lambda_0 \subset T^*{\bf R}^n_x$ the Bessel cylinder (\ref{BC}). Since $p\,dx=d\varphi$ on
 $\Lambda_0$ we get that $(\varphi,\psi)$ are eikonal coordinates.

Let  $L=\Lambda\subset T^*{\bf R}^{n+1}_{x,t}$ be the integral manifold in the extended phase space. If $m=1$ then according to Euler
identity
$$\dot{S} = p \dot{x} - H = p \, \partial_p H - H = 0$$
along the trajectories and $(\varphi, \psi, t)$ are eikonal coordinates on $\Lambda$. Let
$$\Phi: {\bf R}^N \times {\bf R}^d_x \to {\bf R}, \quad \theta \in {\bf R}^N$$
be a smooth function such that the $N\times (d+N)$ matrix
$\bigl( \partial^2_{\theta \theta} \Phi, \partial^2_{\theta x} \Phi \bigr)$
 has rank $N$ on the critical set
\begin{equation}
  \label{3.3}
  C_\Phi=\left\{ (\theta, x)\in {\bf R}^d\times {\bf R}^N: \partial_\theta \Phi (\theta, x)=0 \right\}.
  \end{equation}
In other words,
$C_\Phi$ defined by the equations (\ref{3.3}) is a smooth submanifold in the sense of the implicit function theorem.
Then
\begin{equation} \label{3.4}
  \iota_\Phi:C_\Phi \to T^\ast{\bf R}^d_x, \quad (\theta, x) \mapsto  \bigl(x, \partial_x \Phi (\theta, x)\bigr)
  \end{equation}
is a Lagrangian (i.\,e. $\iota_\Phi^\ast d p \wedge d x = 0$) immersion,
$L=\iota_\Phi(C_\Phi)$ is an immersed Lagrangian submanifold, and $\Phi$ is called
a \textit{generating family} or \textit{phase function} of $L$ --- see e.\,g. \cite{AVG}, \cite{Iv}.

It is standard to show that the number $N$ of $\theta$-variables can be reduced to $N \leq d$, and the minimal possible $N$
is equal to the co-rank of the projection $L \to {\bf R}_x^d$.

If the system (\ref{3.3}) is degenerate in the sense of the implicit function theorem then
$L=\iota_\Phi(C_\Phi)$ is a singular (isotropic) submanifold. In particular its dimension can be less than $d$. See e.\,g.
\cite{CdV} for a general discussion.

Let  $d=n$ and $L=\Lambda_0 \subset T^*{\bf R}^n_x$ be Bessel cylinder (\ref{BC}). Following \cite{DoNa},
it is convenient to use among the $\theta$-parameters
a Lagrange multiplier $\lambda$ and get for $\Lambda_0$ the generating family
\begin{equation*}
  \begin{aligned}
\Phi&_0(x, \theta) = \varphi + \lambda \langle \omega(\psi), x - \varphi \omega(\psi) \rangle =
(1-\lambda) \varphi + \lambda \langle \omega(\psi), x  \rangle,\\
&\theta = (\lambda, \varphi, \psi) \in {\bf R}^{n+1}.
  \end{aligned}
  \end{equation*}
Let now
$$p=P(\varphi, \psi, t), \quad x=X(\varphi, \psi, t)$$
be the Hamiltonian trajectory with an initial condition on the Bessel cylinder (\ref{BC}):
$$P(\varphi, \psi, 0) = \omega(\psi), \quad X(\varphi, \psi, 0) = \varphi \omega(\psi)$$
We apply then (\ref{3.4}) to $d=n+1$ and $L=\Lambda$ being the flow out of $\Lambda_0$ by $v_H$ in $T^\ast{\bf R}^{n+1}$.\\

\begin{prop}
\label{prop2}
Let $H(x,p)$ be positively
homogeneous of degree 1 with respect to $p$.
Then
\begin{equation*}
\Phi(\theta, x, t) = \varphi + \lambda \langle P(\varphi, \psi, t) , x - X(\varphi, \psi, t) \rangle, \
\theta = (\lambda, \varphi, \psi) \in {\bf R}^{n+1}
\end{equation*}
is a generating family for
$\Lambda \subset T^\ast {\bf R}^{n+1}_{x,t}$
at the points satisfying the inequality $\det(P,P_\psi)\neq0$,
which holds at least for small $t$ (because $\det(P,\partial_\psi P)=1$ for $t=0$). In particular, $\Phi$ satisfies
the initial condition $\Phi(\theta, 0, x)=\Phi_0(\theta,x)$.
\end{prop}

Note that $\Phi$ is the 1-jet on $\Lambda$
of the solution of the Hamilton-Jacobi equation
$$\partial_t\Psi+H(x,\partial_x\Psi)=0, \quad  \Psi|_{t=0}=\langle x,\omega(\psi)\rangle$$
defining $v_h$ in (\ref{v}), namely
$\partial_t\Phi+H(x,\partial_x\Phi) = 0$ after the substitution $\lambda = 1$, $x = X(\varphi, \psi, t)$.

Let $z_0 \in \Lambda_0$ be a non-glancing point. It means that $z_0$ is not a critical point of the restriction $H |_{\Lambda_0}$.

\begin{prop}
\label{prop3}
If $H(x,p)$  is positively
homogeneous of degree 1 with respect to $p$ and $E=H(z_0)$ then
\begin{equation*}
\Phi_+^E(\theta_+, x) = \Phi (\lambda, \varphi, \psi, x, t) + E t, \
\theta_+ = (\lambda, \varphi, \psi, t) \in {\bf R}^{n+2}
\end{equation*}
is a generating family for
$\Lambda_+^E \subset T^\ast {\bf R}^{n}_{x}$
at the points that are close to the Hamiltonian trajectory starting at $z_0$ and satisfy the inequality $\det(P,P_\psi)\neq0$,
which holds at least for small $t$.
\end{prop}

\section{Invariant density using the eikonal coordinates}
\setcounter{equation}{0}

Let $y=(y_1,\cdots,y_d)$ be some local coordinates on $C_\Phi$ extended locally to smooth functions
on ${\bf R}^d\times{\bf R}^N$. Then the non vanishing real function

\begin{equation}\label{3.8}
  \begin{aligned}
    F&[\Phi,dy]={dy\wedge d(\partial_{\theta} \Phi)\over dx\wedge d\theta}=\\
    &{dy_1 \wedge \dots \wedge d y_d \wedge d(\partial_{\theta_1} \Phi)\wedge \cdots\wedge d (\partial_{\theta_N} \Phi) \over
       d x_1 \wedge \dots \wedge d x_d \wedge d\theta_1\wedge\cdots\wedge d\theta_N}
  \end{aligned}
  \end{equation}
is well-defined near $C_\Phi$ as the quotient of two volume forms.

The restriction of this function to $C_\Phi$ is important for computations of the Maslov canonical operator on a Lagrangian submanifold
$L = \iota(C_\Phi)$ via its generating family $\Phi$ --- see e.\,g. \cite{DoMaNaTu}, \cite{DoNaSh}.

\begin{prop} 
  Let $H(x,p)$ be positively homogeneous of degree 1 with respect to $p$, $\Phi$ be the generating family
  for $\Lambda\subset T^*{\bf R}^{n+1}_{x,t}$ from Proposition \ref{prop2}, $\theta=(\lambda,\varphi,\psi)$, and $y=(\varphi,\psi,t)$.
Then
\begin{equation*}
F[\Phi,dy]|_{C_\Phi}=
\pm\det(P,\partial_\psi P).
\end{equation*}
\end{prop}

\section{Specific results near a glancing point}
\setcounter{equation}{0}

If $\Lambda_+^E$ is an immersed Lagrangian submanifold we can compute the contribution of the critical points $t \in (0, + \infty)$
in the integral (\ref{int}) using the Maslov canonical operator on $\Lambda_+^E$ (see \cite{AnDoNaRo3}).
At glancing points $\Lambda_+^E$ can become singular and this procedure does not work.

  Here we consider glancing points which are non-degenerate critical points of $H|_{\Lambda_0}$ in the case $n=2$. We use
  $(x,y,p_x,p_y)$ instead of $(x_1,x_2,p_1,p_2)$, let
$$
\begin{aligned}
\pi_t & : (p_x,p_y,E,x,y,t) \mapsto (x,y,t),
\\
\pi_E & : (p_x,p_y,E,x,y,t) \mapsto (x,y,E)
\end{aligned}
$$
be the natural projections.

\begin{theo}\label{thm1}
  Let $z_0 \in \Lambda_0$ be a glancing point being a non-degenerate minimum or maximum of $H|_{\Lambda_0}$,
$z\in\Lambda$ be a
  point of the trajectory starting at $z_0 \in \Lambda_0$ and the point $\pi_t(z)$
  does not belong to the caustic of $\Lambda$. Then in a neighbourhood of the point $z$ there exists a canonical map
  $$
  \bigl(p_x,p_y,E,x,y,t) \mapsto \bigl( p_{\xi},  p_{\eta}, \e ,  \xi,  \eta,  \tau \bigr),
  $$
  i.\,e.
  \begin{equation*}
  dp_{\xi}\wedge d\xi +  dp_{\eta}\wedge\,d\eta- d\e \wedge d\tau =
  dp_{x}\wedge dx + dp_{y}\wedge dy-  {d E} \wedge dt,
  \end{equation*}
  such that

\noindent $\bullet$ $\pi_E(p_{\xi}, p_{\eta}, \e , \xi, \eta, \tau) = (\xi,\eta,\e )$;

\noindent $\bullet$ $\e $ depends only on $E$ and does not depend on the other coordinates $(p_x,p_y,x,y,t)$;

\noindent $\bullet$ the Lagrangian submanifold $\Lambda$ is defined by the following formulas
\begin{gather*}
S(\xi,\eta,\tau) =  -(\tau^3/3 +  \xi^2 \tau), \quad p_\xi = \partial_\xi S = - 2 \xi \tau,
\\
p_\eta = \partial_\eta S = 0, \quad \e  = - \partial_{\tau} S = \tau^2 + \xi^2.
\end{gather*}
\end{theo}

A proof of this theorem with the help of standard methods of Singularity Theory is going to be published later.

Let us consider the following simplest example. If
\begin{gather*}
H = p_y, \quad S_0(x,y) = x^2 y  + y^3/3, \quad z_0=0,
\\
\Lambda_0= \left\{ p_x = \partial_x S_0 = 2 x y, \; p_y = \partial_y S_0 = x^2 + y^2  \right\}.
\end{gather*}
Then
\begin{gather*}
S(x,y,t) = x^2 (y-t)  + (y-t)^3/3,
\\
\Lambda= \left\{ p_x = \partial_x S, \; p_y = \partial_y S, \; E = - \partial_t S  \right\}.
\end{gather*}
The trajectory starting at $z_0$ is  $p_x=p_y=x=0$, $y=t$, let $z$ be its point for $t=T$. Then the canonical transformation
\begin{gather*}
p_\xi = p_x, \quad p_\eta = p_y - E, \quad \e = E,
\\
\xi = x, \quad \eta=y-T, \quad \tau = t - y
\end{gather*}
gives $S = - x^2 \tau  - \tau^3/3$ according to Theorem \ref{thm1}.

Theorem \ref{thm1} implies the following: for $z \in \Lambda$ as above,
if $U_z \subset \Lambda$ is a sufficiently small neighbourhood of $z$,
then for any smooth function
$b :\Lambda\to {\bf R}$ vanishing outside of $U_z$, there exists smooth function $S, B: {\bf R}^3_{x,y,t} \to{\bf R}$
vanishing outside of $\pi_t(U_z)$ such that
\begin{gather*}
    \Lambda \cap U_z =\{ p_x = \partial_x S, \; p_y = \partial_y S, \; E = - \partial_t S\} \cap U_z,
    \\
    [\mathcal{K}_\Lambda b](x,y,t) = B (x,y,t) \exp{\left\{ \frac{i S(x,y,t)}{h} \right\}}.
\end{gather*}
Therefore the integral (\ref{int}) is rewritten in terms of the Maslov canonical operator on the same $\Lambda$ but in the space $T^\ast {\bf R}^3_{x,y,E}$:
\begin{equation*}
  \int\limits_{0}^{+\infty} [\mathcal{K}_\Lambda b](x,y,t) \exp \left\{ \frac{i E t}{h} \right\} \, d t =
  \int\limits_{-\infty}^{+\infty}  B (x,y,t) \exp \left\{ \frac{i ( S(x,y,t) + E t )}{h} \right\} \, d t.
\end{equation*}
and we can express this value in the coordinates of Theorem \ref{thm1} like in \cite{DoNa1},\cite{NaTo}.

\section{The transition of $\Lambda_+^E$ and its phase}
\setcounter{equation}{0}

Now we describe what happens with the singular Lagrangian submanifold $\Lambda_+^E$ as $\sigma=E-E_0$ changes its sign.
More precisely, let us consider a submanifold
  \begin{equation*}
    \mathop{\mathrm{loc}} \Lambda_+^E=\{ z \in T^*{\bf R}^n : \exists \, t \geq 0,
     z \in \exp tv_{H}( \Lambda_0 \cap U_{z_0} \cap \Sigma_E ) \} \subset \Lambda_+^E
  \end{equation*}
where $U_{z_0}$ is a sufficiently small neighbourhood of the initial point $z_0 \in \Lambda_0$ from Theorem \ref{thm1}.

\begin{cor}
In a neighbourhood of the point $\pi_E(z)$ there exist new local coordinates
$$\bigl( \xi(x,y,E),\eta(x,y,E),\e (E) \bigr), \quad \e (E_0)=0$$
such that the singular La\-g\-ran\-gian submanifold $\loc \Lambda_+^E$ is defined by the following formulas
\begin{gather*}
S(\xi,\eta,\tau) = - (\tau^3/3 +  \xi^2 \tau), \quad p_\xi = \partial_\xi S = - 2 \xi \tau,
\\
p_\eta = \partial_\eta S= 0, \quad \e  = - \partial_{\tau} S = \tau^2 + \xi^2.
\end{gather*}
\end{cor}

\begin{figure}
\begin{center}
\includegraphics[width=7cm, height=4cm]{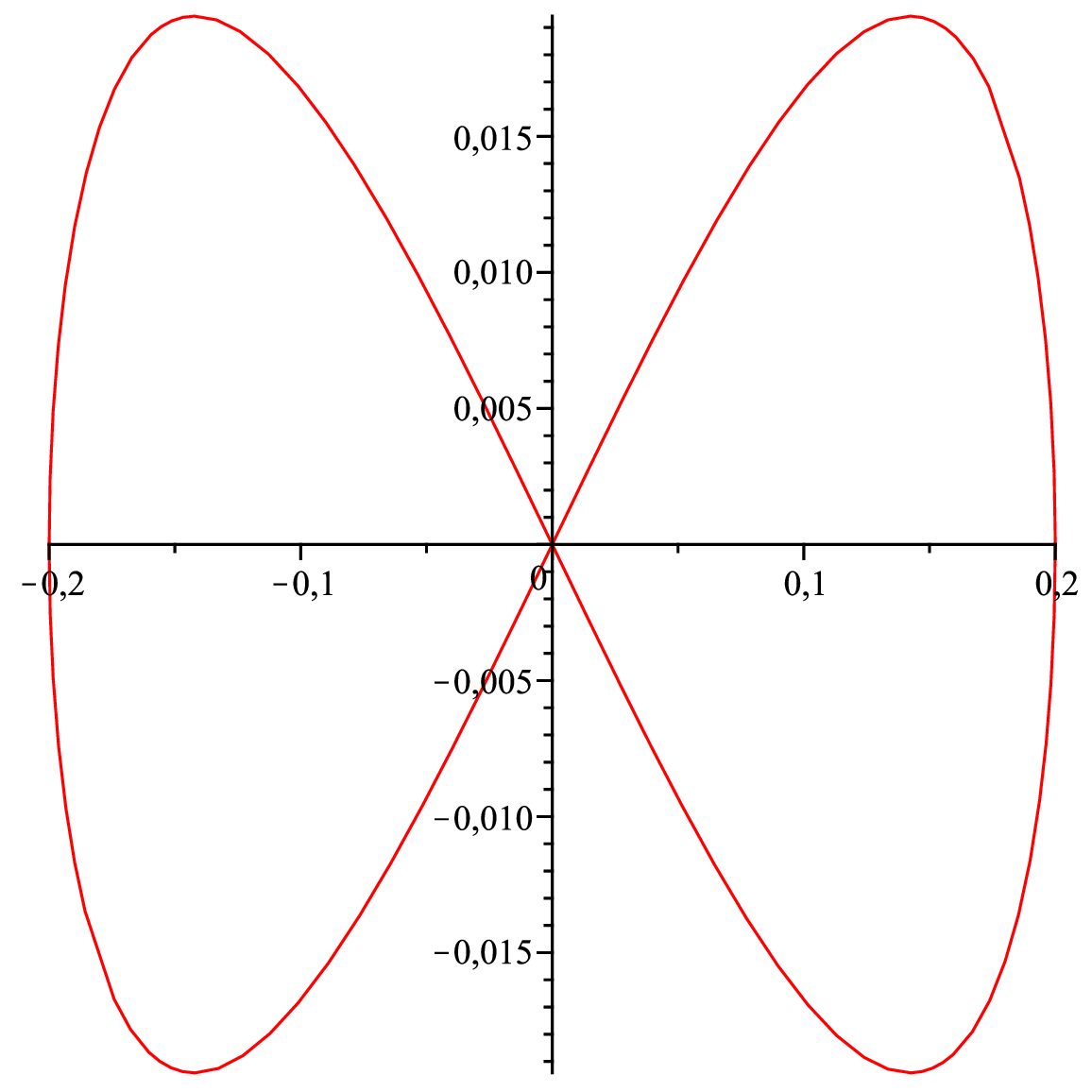}.
\end{center}
\caption{The Lagrangian curve ``$\infty$''}
\label{f1}
\end{figure}

\begin{figure}
\begin{center}
\includegraphics[width=7cm, height=4cm]{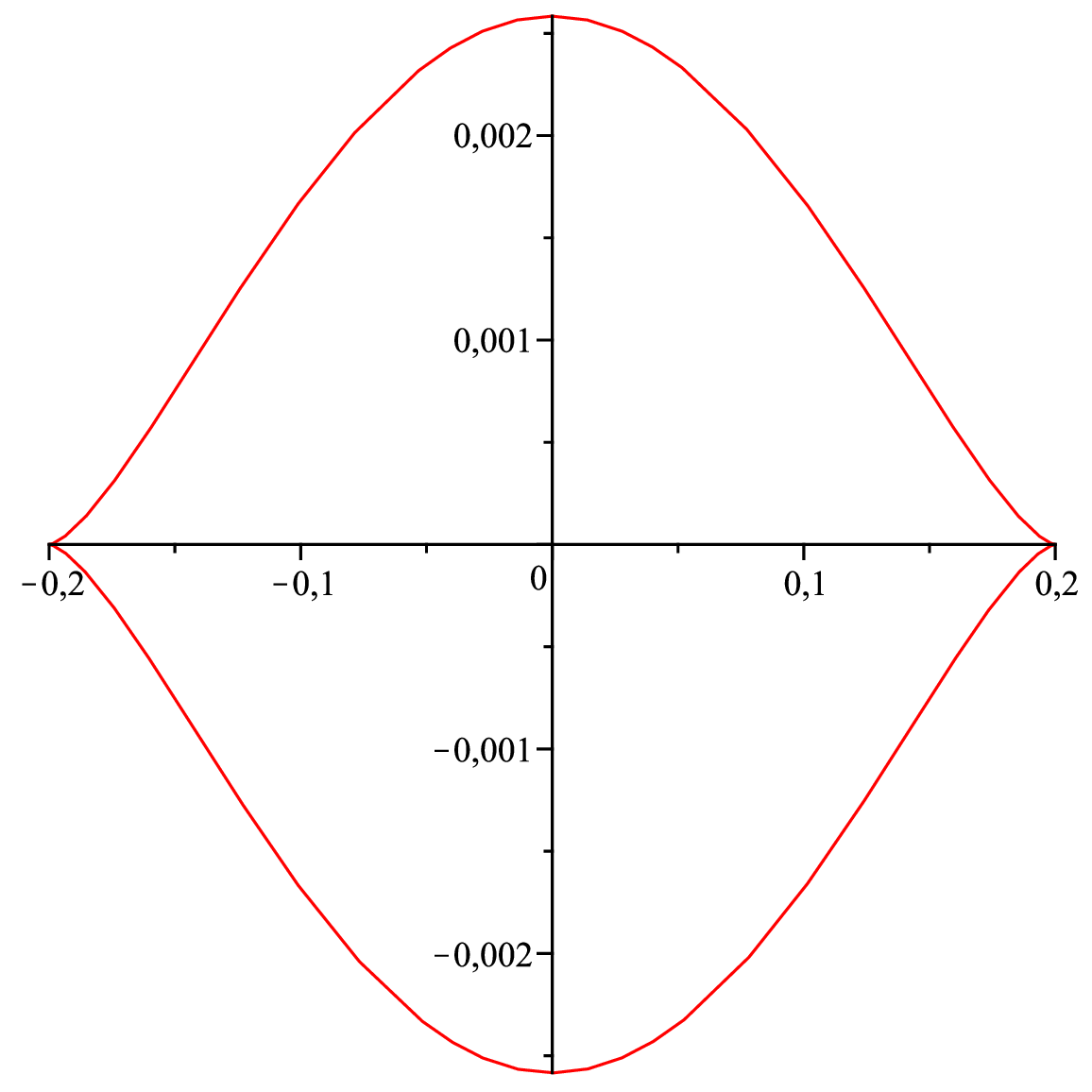}.
\end{center}
\caption{A section of the phase graph}
\label{f2}
\end{figure}

The Lagrangian submanifold $\loc \Lambda_+^E$ is empty if $\e (E) < 0$.
If $\e(E) > 0$ then $\mathop{\mathrm{loc}} \Lambda_+^E$ becomes the cylinder over  the Lagrangian curve ``$\infty$'' shown in Fig.\,\ref{f1}.
The graph of the phase in this case is  the cylinder over the curve with two cusps shown in Fig.\,\ref{f2}.
If $\e(E) = 0$ then $\mathop{\mathrm{loc}} \Lambda_+^E$ degenerates into an interval of the Hamiltonian phase trajectory starting at $z_0$.

The above transition is realized, for example, if
$H=(p_x^2+p_y^2)/{\rho(x,y)}$,
$\rho$ has a non-degenerate maximum or minimum at the origin, and
$f_h = J_0 \bigl( \sqrt{(x-a)^2+y^2}/{h} \bigr), \quad a \neq 0$.
In this case $\Lambda_0$ is the cylinder $p_x = \cos{\phi}$, $p_y = \sin{\phi}$, $x = a + \tau \cos{\phi}$, $y = \tau \sin{\phi}$.
As the ``source'' $(a,0)$ goes to $- \infty$ $f_h$ becomes the simple wave
$\exp{\frac{i x}{h}}$
and $\Lambda_0$ is the plane $p_x=1$, $p_y=0$. Here is the simplest example
$f_h = \exp{\frac{i x}{h}}, \quad \varrho = 1 + x^2 + y^2$
where one can get all explicit formulas. Our first figure is the dependence between $y$ (the horizontal axe) and $p_y$ (the vertical axe) if $x=0$.
The second figure is the dependence between $y$ (the horizontal axe) and the phase (the vertical axe). The first curve is the derivative of the second one.

\section{Glancing hypersurfaces and La\-g\-ran\-gi\-an intersections}
\setcounter{equation}{0}

We discuss here a possible application of non transverse Lagrangian intersections to problems of diffraction by an obstacle.
Let $M$ be a smooth manifold, since we are working locally, we will assume $M={\bf R}^n$.
Let $F,G$ be two smooth hypersurfaces of $T^*M$,  intersecting
transversally at $z$. Recall \cite[Definition 21.4.6]{Ho} that $F$ and $G$ are said to be {\it glancing} at $z$ iff
the Hamilton foliation of $F=\{f=0\}$ and
$G=\{g=0\}$ (locally near $z$) are simply tangent at $z$.

Stated otherwise, we have $f(z)=g(z)=\{f,g\}(z)=0$ (Poisson bracket),
but the second Poisson brackets $\{f,\{f,g\}\}(z), \{g,\{g,f\}\}(z)$ are non zero.

By the theorem of equivalence of glancing hypersurfaces of Melrose \cite[Theorem 21.4.8]{Ho}
there are local symplectic coordinates $(x,\xi)$ vanishing at $z$ such that
$F,G$ are defined resp. by $x_1=0$ and $\xi_1^2-x_1-\xi_2=0$.
Then $g=\xi_1^2-x_1-\xi_2=0$ will be the ``normal form'' of $H$ in these coordinates. (We use the notation $g$ for the normal form of $H$, or $H-E$).

We apply this theorem to $G$ being the energy surface $H=E$ (i.\,e. $g=0$) and $F$ an auxiliary hypersurface intersecting $G$ transversally at
a glancing point $z$. We want to find some germs of Lagrangian manifolds $\Lambda$ such that $\Lambda$ is transverse to $F$ at $z$
but $(\Lambda,G)$ has glancing intersection at $z$.
This means that $T_z\Lambda\cap(T_zF)^\sigma=\{0\}$ and ${\bf R}v_H(z)=(T_zG)^\sigma\subset T_z\Lambda$, where superscript $\sigma$
denotes symplectic orthogonal.
Assume $n=2$ for simplicity.

Define $\Lambda$ locally near $z$ by $f_1=f_2=0$, with $\{f_i,f_j\}=0$. Consider the symmetric matrix
\begin{equation*}
A_z=A_z(\Lambda,G)=
\begin{pmatrix}\{f_2,\{f_2,g\}\}&-\{f_1,\{f_2,g\}\}\\ -\{f_2,\{f_1,g\}\}&\{f_1,\{f_1,g\}\}\end{pmatrix}(z)
\end{equation*}
and the vector
$$
B_z=B_z(\Lambda,G)={\{g,\{g,f_1\}\}\choose\{g,\{g,f_2\}\}}(z).
$$
Let $z$ be a {\it glancing point} for the pair $(\Lambda,G)$ i.\,e.
\begin{equation*}
g(z)=f_1(z)=f_2(z)=0, \ \{g,f_1\}(z)=\{g,f_2\}(z)=0.
\end{equation*}
We distinguish the following 10 possibilities for the 2-jets of $(\Lambda,G)_z$ (not covering the entire classification of \cite{ZaMy}):
\begin{enumerate}
\item
$\det A_z> 0, \ B_z\neq0$;
\item
$\det A_z > 0, \ B_z = 0$;
\item
$\det A_z < 0, \ {}^tB_z A_z B_z \neq 0$;
\item
$\det A_z < 0, \ {}^tB_z A_z B_z = 0, \ B_z\neq 0$;
\item
$\det A_z < 0, \ B_z = 0$;
\item
$\det A_z = 0, \ {}^tB_z A_z B_z \neq 0$;
\item
$\det A_z = {}^tB_z A_z B_z = 0, \ A_z \neq 0, \ B_z \neq 0$;
\item
$\det A_z = \ B_z = 0, \ A_z \neq 0$;
\item
$A_z = 0, \ B_z \neq 0$;
\item
$A_z = B_z = 0$.
\end{enumerate}

We know that a general Lagrangian manifold can be parametrized in the mixed representation, so when $n=2$ by one of the following cases
\begin{equation*}
  \begin{aligned}
(I)  \; & \Lambda=\left\{p=\partial_x\phi \right\}, \\
(II) \; & \Lambda=\left\{x=-\partial_p\phi \right\}, \\
(III)\; & \Lambda=\left\{x_1=-\partial_{p_1}\phi, p_2=\partial_{x_2}\phi \right\}, \\
(IV) \; & \Lambda=\left\{p_1=\partial_{x_1}\phi, x_2=-\partial_{p_2}\phi \right\}.
  \end{aligned}
  \end{equation*}
We have
\begin{prop}
Assume $\Lambda$ as above to be pa\-ra\-met\-ri\-zed by a quadratic phase $\phi=\phi_0$.
In Cases (I), (II), (III) $\phi_0$ are one-parameter families taking values:

(I) $\phi_0(x)={1\over2}(ax_1^2-2x_1x_2)$;

(II) $\phi_0(\xi)={1\over2}(2\xi_1\xi_2+c\xi_2^2)$;

(III) $\phi_0(x_2,\xi_1)={1\over2}b(\xi_1+x_2)^2$

\noindent respectively, where $a,b,c\neq0$. The manifold $\Lambda$ is transverse to $F$ at $z=0$,
and the corresponding matrices $A_z(\Lambda,G),B_z(\Lambda,G)$ above are then given by:

(I) $A_z=2\begin{pmatrix}1&a\\ a&a^2\end{pmatrix}$, $B_z=2\begin{pmatrix}-a \\ 1\end{pmatrix}$;

(II) $A_z=2\begin{pmatrix}0&0\\ 0&1\end{pmatrix}$, $B_z=2 \begin{pmatrix} 1 \\ 0 \end{pmatrix}$;

(III) $A_z=2\begin{pmatrix}0&0\\ 0&1\end{pmatrix}$, $B_z=2 \begin{pmatrix} 1 \\ 0 \end{pmatrix}$.

\noindent
At last, Case (IV) does not occur.
\end{prop}

\end {document}